\documentclass[aps,prb,reprint,amsmath,amssymb,superscriptaddress,longbibliography]{revtex4-2}

\newcommand{\RCMO}{Rb$_2$Cu$_2$Mo$_3$O$_{12}$\xspace}
\newcommand{\CCMO}{Cs$_2$Cu$_2$Mo$_3$O$_{12}$\xspace}

\usepackage{graphicx}
\usepackage{dcolumn}
\usepackage{bm}
\usepackage{amsmath,amssymb,amsfonts,mathrsfs}
\usepackage{xcolor}
\usepackage{xspace}
\usepackage{xfrac}
\usepackage{color,soul}
\usepackage{graphicx}
\usepackage[flushleft]{threeparttable}
\usepackage{multirow}
\usepackage{booktabs}
\usepackage[colorlinks=true,urlcolor=blue,citecolor=blue,linkcolor=blue,bookmarks=false,pdfstartview={FitH}]{hyperref}

\begin{document}
	
	\title{Dielectric relaxation in the quantum multiferroics \RCMO and \CCMO}
	
	\author{D. Flavi\'{a}n}
	\affiliation{Laboratory for Solid State Physics, ETH Z{\"u}rich, 8093 Z{\"u}rich, Switzerland}
	
	\author{P.  A. Volkov}
	\affiliation{Department of Physics, University of Connecticut, Storrs, Connecticut 06269, USA}
	
	\author{S.  Hayashida}
	\affiliation{Laboratory for Solid State Physics, ETH Z{\"u}rich, 8093 Z{\"u}rich, Switzerland}
	\affiliation{Neutron Science and Technology Center, Comprehensive Research Organization for Science and Society (CROSS), Tokai, Ibaraki 319-1106, Japan}
	
	\author{K.~Yu.~Povarov}
	\affiliation{Dresden High Magnetic Field Laboratory (HLD-EMFL) and W\"urzburg-Dresden Cluster of Excellence ct.qmat, Helmholtz-Zentrum Dresden-Rossendorf, 01328 Dresden, Germany}	
	
	\author{S.  Gvasaliya}
	\affiliation{Laboratory for Solid State Physics, ETH Z{\"u}rich, 8093 Z{\"u}rich, Switzerland}
		
	\author{P.  Chandra}		
	\affiliation{Department of Physics and Astronomy, Center for Materials Theory, Rutgers University, Piscataway, NJ 08854, USA}
		
	\author{A. Zheludev}
	\affiliation{Laboratory for Solid State Physics, ETH Z{\"u}rich, 8093 Z{\"u}rich, Switzerland}
	
	\begin{abstract}
	
		Motivated by the recent discovery of dielectric relaxation by quantum critical magnons in \CCMO, we conduct a detailed analysis of its dielectric response and compare it to that in the isostructural compound \RCMO. Measurements in the vicinity of the field-induced magnon softening show that its description in terms of 3D Bose-Einstein condensation of magnons quantum critical point is unaltered by the presence of dielectric relaxation. We also demonstrate the existence of dielectric relaxation anomalies at 19 K in \RCMO and discuss the implications for the microscopic origin of dielectric activity in two compounds.

	\end{abstract}
	\maketitle

In our previous work we reported unusual dielectric properties at magnetic quantum phase transitions in two isostructural frustrated quantum antiferromagnets, namely \RCMO \cite{hayashida2021} and \CCMO \cite{Flavian130Relaxation}. Thanks to magneto-electric coupling, the magnetically ordered phases of \RCMO are also ferroelectric under a finite magnetic field. In applied fields, the electric polarization becomes directly coupled to the antiferromagnetic order parameter. As a result, dielectric permittivity shows a critical divergence at the magnetic phase transitions. This includes the magnetic field-induced saturation transition and the corresponding quantum critical point (QCP). Since the latter can be described in terms of Bose-Einstein condensation (BEC) of magnons 	\cite{Batyev1984,Giamarchi1999,Nikuni2000BEC,Giamarchi2008,Zapf2014}, dielectric experiments enabled a direct excitation-probe measurement of the critical susceptibility and the critical exponent $\gamma$ \cite{hayashida2021}. This is significant, as in conventional BEC those quantities are not physically accessible.  

An apparently similar behavior of the dielectric constant at magnetic phase boundaries was also detected in \CCMO. However, the focus of previous work was on an additional new dielectric anomaly, one that is not seen in the Rb-based system \cite{Flavian130Relaxation}. This broad anomaly does not explicitly follow the magnetic phase boundary, but nevertheless shows critical characteristics at the QCP. Its origin is entirely different.  It represents Debye-like relaxation of some intrinsic dipolar degrees of freedom mediated by quantum-critical soft magnons. This demonstrates that magnetic excitations in this system show electric activity of their own, regardless the existence of long-ranged ordering.

\begin{figure}[h]
		\centering
		\includegraphics[scale=1]{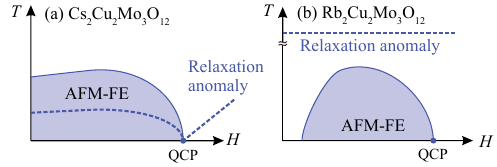}
		\caption{Schematic phase diagrams of (a) \CCMO and (b) \RCMO showing their antiferromagnetic (AFM) and ferroelectric (FE) long range order domes (shaded) and the presence of relaxation anomalies (dashed line) as discussed in detail in the text.}
		\label{fig:Summary}
\end{figure}

Several important questions remain unanswered.  Is the sharp dielectric anomaly at the phase boundaries in \CCMO also described in terms of critical BEC susceptibility, as in \RCMO? How is the relaxation anomaly in \CCMO affected by the spin-flop transition, which is present in that material but absent in the Rb-counterpart?  What is the origin of the dipolar degrees of freedom in \CCMO and are they also present in \RCMO? And if they are ineed also present there, why does the relaxation anomaly seem to be absent? In the present work we address these issues via new dielectric measurements on the Cs-compound, highlighting similarities and differences between the materials (Fig.\ref{fig:Summary}).
	
	\section{Results}
	
\begin{figure}[tbp]
	\centering
	\includegraphics[scale=1]{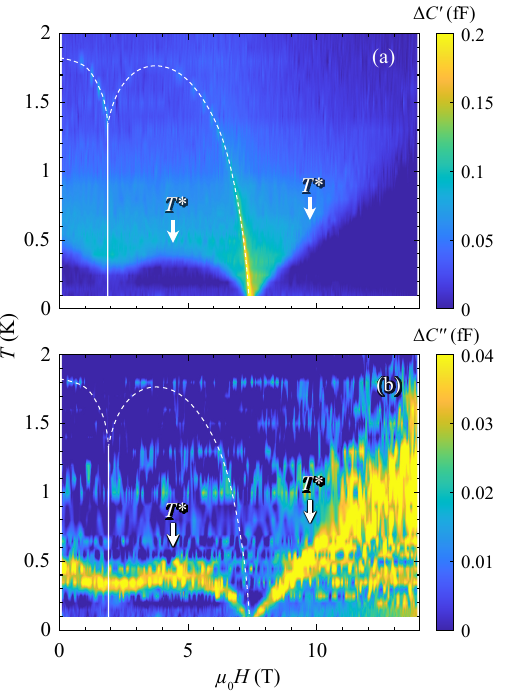}
	\caption{False color plots of complex capacitance $\Delta C = \Delta C^\prime+i \Delta C^{\prime\prime}$ in \CCMO : (a) real and (b) imaginary parts. The electric field is applied parallel to the \textit{c}$^*$ axis, and magnetic field along the \textit{b} axis. A dashed line shows the phase boundary as extracted from specific heat measurements in Ref.~\cite{Flavian2020}, while a solid line represents the spin-flop field. Arrows indicate the relaxing feature at $T^\ast$ in both real and imaginary plots.}
	\label{fig:Capacitance_B}
\end{figure}
	
	\subsection{Capacitance for $\mathbf{H}\|\mathbf{b}$}

	In the present study we used the same 2.4$\times$0.25$\times$0.18 mm$^3$, and 0.25~mg single crystal \CCMO sample and experimental setup as in Ref.~\cite{Flavian130Relaxation}.  Sample capacitance is measured with an alternating current (AC) probing field at 1~kHz, as are all subsequent measurements in this text. Data for $\mathbf{H}\|\mathbf{b}$ and $\mathbf{E}\|\mathbf{c}^\ast$ is presented in Fig.~\ref{fig:Capacitance_B} as a function of magnetic field and temperature. False color plots show the real ($\Delta C^\prime$) and imaginary ($\Delta C^{\prime\prime}$) components of the complex signal. These plots are to be compared against Fig.~1 in Ref.~\cite{Flavian130Relaxation}, which shows similar data for other magnetic field directions. The same anomalies are also found here: a sharp feature at the phase boundary as a result of magnetic/ferroelectric order, and a relaxing feature with significant real and imaginary parts. As previously, the temperature at which the imaginatry part of the latter is a maximum  is denoted as $T^\ast$.

The presence of a spin-flop transition at 2 T for this field configuration notably affects both kinds of anomalies.  On the one hand,  the feature at the phase boundary retains a significant weight as the magnetic field is lowered almost all the way down to zero field. It is particularly prominent in the vicinity of the bicritical termination point found at 2 T and 1.4 K.  On the other hand, the value of $T^*$ is visibly reduced around the spin-flop field, echoing the reduction of the ordering temperature, $T_c$. 
	
	\subsection{Effect of a bias electric field}
	
\begin{figure}[tbp]
	\centering
	\includegraphics[scale=1]{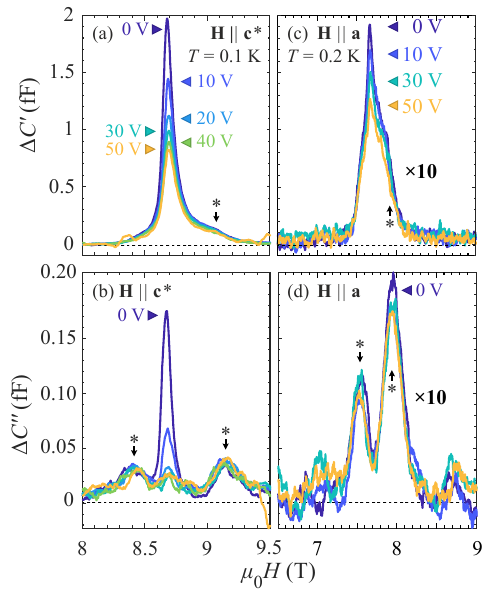}
	\caption{Effect of a DC bias electric field along the crystallographic \textit{c}$^*$ axis on the complex dielectric response of \CCMO close to the QCP for: (a,b) \textbf{H}$\parallel$\textbf{c}$^\ast$ and (c,d) \textbf{H}$\parallel$\textbf{a}.  Note that data in (c,d) have been multiplied by 10 in order to match the scale in (a,b). Colors show field scans at a constant temperature of and different bias fields from 0 to $\sim$250 kV/m. Only the diverging feature at the phase boundary is modified by a uniform electric field, as expected for a critical susceptibility.  The broad relaxing features, shown by asterisks in (a-d), are unchanged.}
		\label{fig:BiasField}
\end{figure}	
	
	For \RCMO the fact that the dielectric polarization is a primary order parameter of the magnetic phase transitions was established by studying the effect of a static electric field.  Under such an applied field, the divergence of dielectric permittivity at the phase boundary was shown to be suppressed, and the corresponding peak broadened. We now conducted the same measurement in \CCMO to understand how a static field affects both the phase boundary and relaxation anomalies. In all cases the direct current (DC) electric field was applied parallel to the probing AC field. Figure~\ref{fig:BiasField} shows data obtained for two orthogonal field directions, where the boundary signal is strongest ($\mathbf{H}\|\mathbf{c}^*$) and weakest ($\mathbf{H}\|\mathbf{a}$).

\begin{figure}[tbp]
	\centering
	\includegraphics[scale=1]{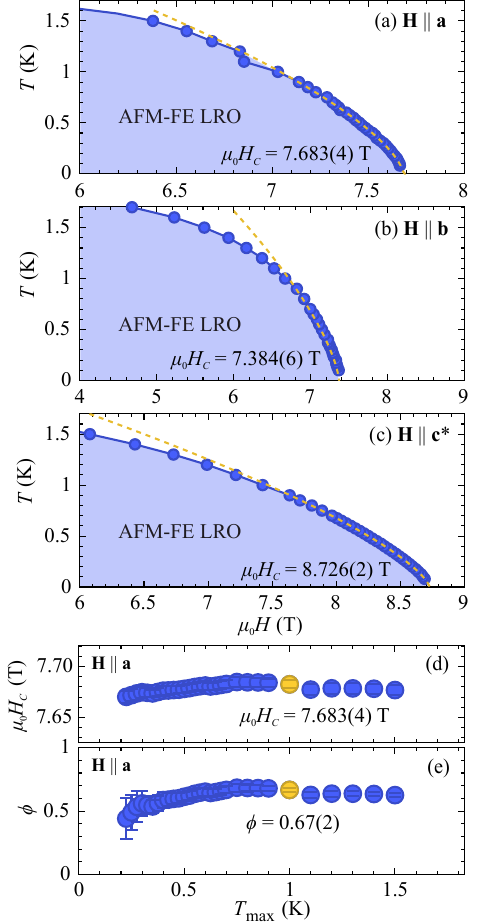}
	\caption{Analysis of the phase boundaries of antiferro- magnetic-ferroelectric (AFM-FE) long-range order (LRO) dome in \CCMO. Magnetic field is applied along the crystallographic directions (a) \textit{{a}}, (b) \textit{{b}},  and (c) \textit{{c}}$^*$. Circles show the phase boundary extracted from typical constant-temperature field scans.  The shaded area shows the long-range-ordered region.  A dashed line is the best fit to a power-law dependence. The best fits (shown in Table~\ref{table:Exponents}) are obtained in a windowing analysis, as exemplified in (d) and (e) for \textbf{H}$\parallel$\textbf{a}.  $T_\mathrm{max}$ in the abscissa represents the maximum temperature of the window used in each individual fit. The boundary exponent is in perfect agreement with BEC criticality ($\phi=2/3$). 
	}
	\label{fig:BECScaling}
\end{figure}		
	
	The suppression of the phase boundary anomaly with an increasing electric field is fully analogous to what was seen in \RCMO. Note that the effect on the imaginary part is particularly strong. This component of dielectric permittivity is associated with ferroelectric domain-wall dynamics \cite{liu2015losses}. A static field suppresses domain formation and thereby the imaginary part of the response. 
	
	In contrast, the behavior of the relaxing features is completely different. The external field has virtually no effect on the signal at fields close to the QCP for either magnetic field orientation. The relaxing features in the real part of capacitance, that appear as a broad plateau of about 0.5~T in width below the divergent peak [as shown by asterisks in Figs.~\ref{fig:BiasField}(a) and \ref{fig:BiasField}(c)], are not altered by fields as high as 250~kV/m (50 V). Similarly, the imaginary part [asterisks Figs.~\ref{fig:BiasField}(b) and \ref{fig:BiasField}(d)] remains unchanged over the whole range of electric fields. This observation confirms that the relaxation anomaly is entirely unrelated to ferroelectric long-range order in the system.
	
	\subsection{Shift exponents}

The very sharp and prominent dielectric anomaly at the phase boundary allows a very precise measurement of the field dependence of the ordering temperature, and with it the shift exponent $\phi$ at the magnetic saturation QCP. The positions of the dielectric peaks are plotted versus field and temperature in Fig.~\ref{fig:BECScaling} for three different applied field directions. In order to determine both the critical field and the shift exponents for the three studied geometries, we employ a shrinking-window fit analysis \cite{Huvonen2012}, as was previously done for \RCMO. This procedure makes the best of the experimental data while trying to restrict the analysis on a region as narrow as possible around the critical point. The critical field $H_c$, the power law exponent $\phi$ and a scale factor are used as parameters in a least-squares fitting, as a function of temperature window given by $T_\mathrm{max}$. The optimal fitting window is narrowed until a further reduction stops producing a statistically significant change in the fit results. An example of such analysis is provided in Figs.~\ref{fig:BECScaling}(d) and \ref{fig:BECScaling}(e).  The best fits from the windowing analysis are shown in Figs.~\ref{fig:BECScaling}(a)-\ref{fig:BECScaling}(c) and summarized in Table~\ref{table:Exponents}. 

The fitted shift exponents are in good agreement with the expected $\phi = 2/3$ for a 3D BEC QCP \cite{Batyev1984,Giamarchi1999,Nikuni2000BEC,Giamarchi2008,Zapf2014}. The present case reveals a much better agreement with the BEC framework than \RCMO, where the presence of a narrow presaturation phase for certain field orientations is associated with a distorted phase boundary. Even though the presence of presaturation phases in \CCMO has been found for all field orientations \cite{Flavian2020}, these seem to have a minimal effect on the shape of the long-range-ordered dome. 

\begin{table}[tbp]
\centering
\caption{Critical fields and boundary exponents at the QCP of \CCMO. Values are extracted from best fits in the range given by $T_\mathrm{max}$ around the QCP for the investigated field geometries. }
\label{table:Exponents}
\begin{tabular}{c|c|c|c}
 \midrule  \midrule
&  \;\;\;\; $\mu_0 H_c$ (T) \;\;\;\; &\;\;\;\;\;\; $\phi$\;\; \;\;\;\; & \;\; \;$T_\mathrm{max}$ (K) \;\; \;\\ 
\cmidrule( l r){1-4}
\cmidrule( l r){1-4}
\;\; $\textbf{H}\parallel\textbf{a}$ \;\; & 7.683(4) & 0.67(2) & 1.00 \\ 
\cmidrule( l r){1-4}
\;\; $\textbf{H}\parallel\textbf{b}$\;\; & 7.384(4) & 0.68(3)  & 0.90 \\ 
\cmidrule( l r){1-4}
\;\; $\textbf{H}\parallel \textbf{c}^\ast$\;\; & 8.726(5) & 0.712(5)  & 0.90\\ 
 \midrule  \midrule
 \end{tabular} 
\end{table}

\subsection{Critical susceptibility}

\begin{figure}[tbp]
	\centering
	\includegraphics[scale=1]{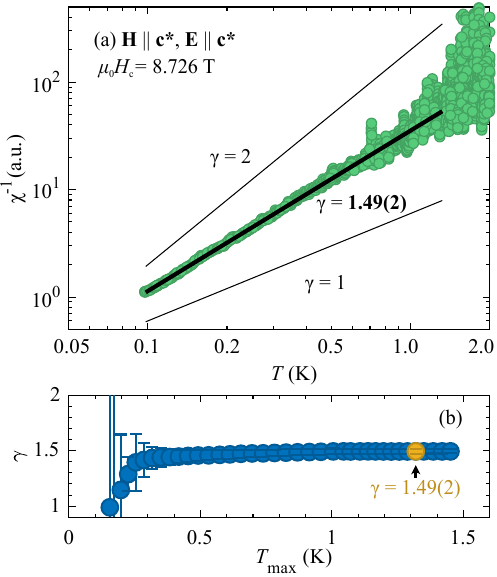}
	\caption{(a) Scaling of the critical dielectric susceptibility ($\chi$) of \CCMO along the critical trajectory for \textbf{H}$\parallel$ \textbf{c}$^*$  (fixed $\mu_0H_c = 8.726 $~T), after subtraction of the non-critical contribution from Fig.~\ref{fig:ColeFitCstar}(a).  A double logarithmic plot shows indeed power-law behavior below 2 K and down to the lowest measured temperatures.  Best fit to a power-law trend is shown as a solid thick line, along with the obtained critical exponent. Solid thin lines are guides to the eye representing other typical critical exponents.  (b) Results of a windowing analysis performed to analyze the critical susceptibility. The optimal fitting range is highlighted. }
	\label{fig:QCPCriticality}
\end{figure} 

	Having determined the critical fields we turn to study the BEC critical susceptibility, $\chi$, at the QCP.  It was already shown in \cite{Flavian130Relaxation} that a non-critical contribution to dielectric permittivity exists at low temperature, as a result of the existence of intrinsic electric dipoles in \CCMO. This field-independent contribution ($\Delta C_0$) is analyzed below in detail (see Fig.~\ref{fig:ColeFitCstar}(a)) and is subtracted from the measurement of $\Delta C$ to obtain the critical dielectric susceptibility, $\chi$. Figure~\ref{fig:QCPCriticality}(a) shows precisely this quantity for \textbf{H}$\parallel$ \textbf{c}$^*$ along the critical trajectory with $\mu_0H_c = 8.726$~T. The double logarithmic scale in Figure~\ref{fig:QCPCriticality}(a) shows a robust power-law trend for $\chi$ over a decade in temperature and down to the lowest accessible values. A shrinking window analysis is used to extract a consistent power-law exponent, just as in Fig.~\ref{fig:BECScaling} and is illustrated in Fig.~\ref{fig:QCPCriticality}(b).  Given the high density of data points, these are binned together every 40 points for the analysis. Best fit to a $T^{-\gamma}$ dependence is obtained with $T_\mathrm{max} = 1.31$~K and provides a critical exponent $\gamma = 1.49(2)$, in fantastic agreement with $\gamma = 3/2$ as expected for a three-dimensional BEC QCP \cite{Continentino2011}.  
	
	Just as in the case of \RCMO, the saturation transition in \CCMO is an excellent example of magnon-BEC criticality.  The results in Fig.~\ref{fig:QCPCriticality} show that the BEC description of the transition is unaltered by the presence of dipolar relaxation in \CCMO, and further establishes the independence of ferroelectric ordering and relaxation processes. 

\subsection{Dielectric relaxation by magnons}

\begin{figure}[tbp]
	\centering
	\includegraphics[scale=1]{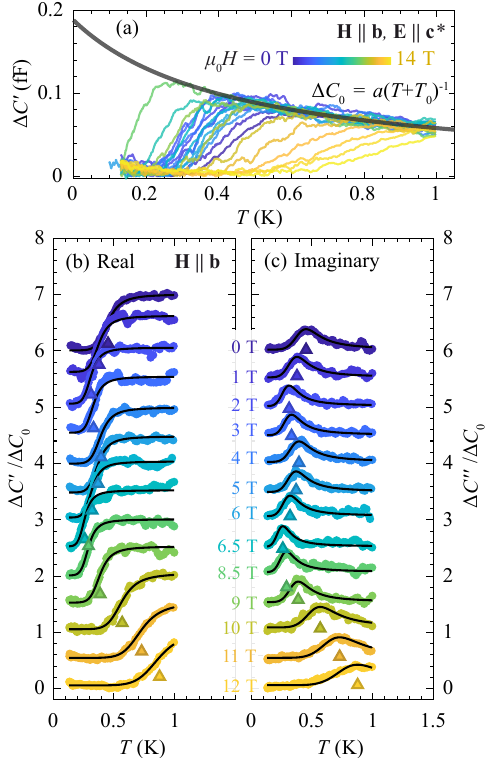}
	\caption{Typical constant-field measurements of complex capacitance of \CCMO for $\textbf{H}\parallel\textbf{b}$ and $\textbf{E}\parallel\textbf{c}^\ast$. (a) Measurements at different fields are shown in colors, showing a perfect collapse at high temperature, from which the static response $\Delta C_0$ is obtained (thick shaded line). (b,c) Capacitance measurements normalized by $\Delta C_0$: (b) real, and (c) imaginary components. Black solid lines show a fit to \eqref{eq:Cole-Cole}. Triangles show the value of $T^\ast$ as defined in the text. Each data set is offset vertical by 0.5 units for visibility. }
	\label{fig:ColeFitB}
\end{figure} 

\begin{figure}[tbp]
	\centering
	\includegraphics[scale=1]{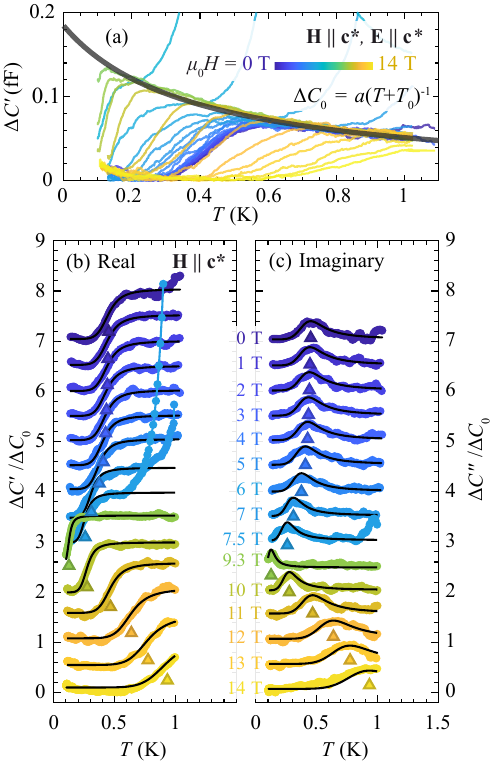}
	\caption{Typical constant-field measurements of complex capacitance of \CCMO for $\textbf{H}\parallel\textbf{c}^\ast$ and $\textbf{E}\parallel\textbf{c}^\ast$. (a) Measurements at different fields are shown in colors, showing a perfect collapse at high temperature, from which the static response $\Delta C_0$ is obtained (thick shaded line). (b,c) Capacitance measurements normalized by $\Delta C_0$: (b) real, and (c) imaginary components. Black solid lines show a fit to \eqref{eq:Cole-Cole}. Triangles show the value of $T^\ast$ as defined in the text. Each data set is offset vertical by 0.5 units for visibility.  The prominence of the boundary dielectric anomaly close to the QCP is apparent in this geometry.}
	\label{fig:ColeFitCstar}
\end{figure} 

We extend the analysis presented in Ref.~\cite{Flavian130Relaxation} to a wider set of dielectric data, including two additional orientations of the external magnetic field. The magnon-mediated relaxation of electric dipoles manifests itself in a Cole-Cole relaxation process \cite{debye1929polar,Cole-Cole}, modeled after the following dependence:
\begin{equation}
\Delta C = \frac{\Delta C_0 (T)}{1+[i\omega\tau]^{1-\alpha}}
\label{eq:Cole-Cole}
\end{equation}
where best agreement is found for activated behavior of the relaxation time $\tau(T,H) = \tau_0 \exp[E_b(H)/k_BT]$.  Fig.~\ref{fig:ColeFitB} and Fig.~\ref{fig:ColeFitCstar} show capacitance data for the geometries with $\textbf{H}\parallel\textbf{b}$ and $\textbf{H}\parallel\textbf{c}^\ast$, respectively.  All the data are analyzed following the prescriptions in \cite{Flavian130Relaxation}. A field-independent static response, $\Delta C_0$, is obtained for every geometry, with $T_{0, \textbf{b}}= 0.44(7)$~K and $T_{0, \textbf{c}^\ast}= 0.39(2)$~K, as shown in Fig.~\ref{fig:ColeFitB}(a) and Fig.~\ref{fig:ColeFitCstar}(a), and in good agreement with the results reported in \cite{Flavian130Relaxation}. Further, the entirety of the data set is fitted to \eqref{eq:Cole-Cole} using a single parameter $\alpha = 0.2$ and a single attempt time $\tau_0=10^{-6}$~s as global variables, same values as in in the previous report, emphasizing the field independence of these parameters.  Optimization of the activation energy, $E_b(H)$, for each magnetic field value results in a good agreement with the experimental data as shown in Fig.~\ref{fig:ColeFitB}(b,c) for $\textbf{H}\parallel\textbf{b}$, and in Fig.~\ref{fig:ColeFitCstar}(b,c) for $\textbf{H}\parallel\textbf{c}^\ast$.

\begin{figure}[tbp]
		\centering
		\includegraphics[scale=1]{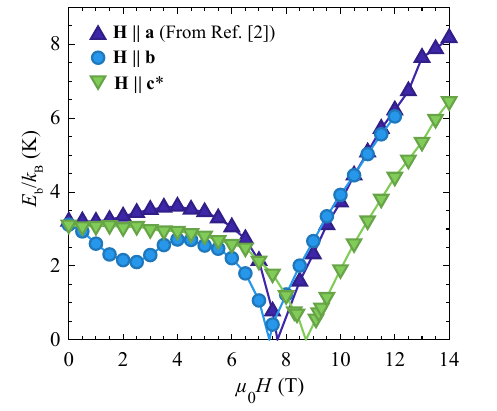}
		\caption{Characteristic energy barriers ($E_b$) in \CCMO extracted from fits of complex capacitance to a Cole-Cole model \eqref{eq:Cole-Cole}.  Different symbols represent magnetic fields along different crystallographic directions; solid lines are a guide to the eye and reach zero at the corresponding critical fields, as given in Table~\ref{table:Exponents}. }
		\label{fig:Relaxation_Magnons}
\end{figure}

The resulting energy barriers for all three studied geometries are displayed in Fig.\ref{fig:Relaxation_Magnons}. As was shown in \cite{Flavian130Relaxation} the energy barrier approaches zero at the QCP. Above $H_c$, the energy barrier follows the Zeeman-induced single-magnon energy gap $E_{b,i} = g_i \mu_B \mu_0 (H-H_{c,i})$, with $g_i$ the $g$-factor and $i = a,  b,  c^*$. Linear fits close to the transition result in $g$-factors $g_\textbf{a} = 2.13$, $g_\textbf{b} = 2.11$ and $g_{\textbf{c}^*}=2.10$, in fantastic agreement with the isotropic $g$-tensor found in bulk measurements \cite{Flavian2020}. 

In addition,  results for $\textbf{H}\parallel\textbf{b}$ evidence the influence of the spin-flop transition in the dipolar relaxation rate. The height of the energy barrier shows a dip at the spin-flop field, the same way $T_c$ shows a dip.  This indicates that the relaxation rate is linked to the underlying magnetic structure and the associated magnetic excitations.  However, we reiterate that long range magnetic order is not a requirement for magnon-mediated relaxation to take place. 

\subsection{Dielectric relaxation in \RCMO}
\begin{figure}[tbp]
	\centering
	\includegraphics[scale=1]{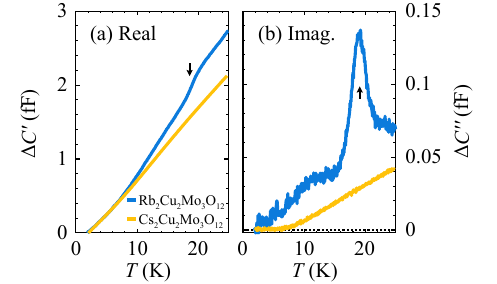}
	\caption{Zero-field complex capacitance on single crystals of \CCMO and \RCMO. A black arrow in (a) real and (b) imaginary panels indicate the presence of a dielectric anomaly in \RCMO, which is completely absent in \CCMO.}
	\label{fig:Comparison}
\end{figure}	

\begin{figure*}[tbp]
		\centering
		\includegraphics[scale=1]{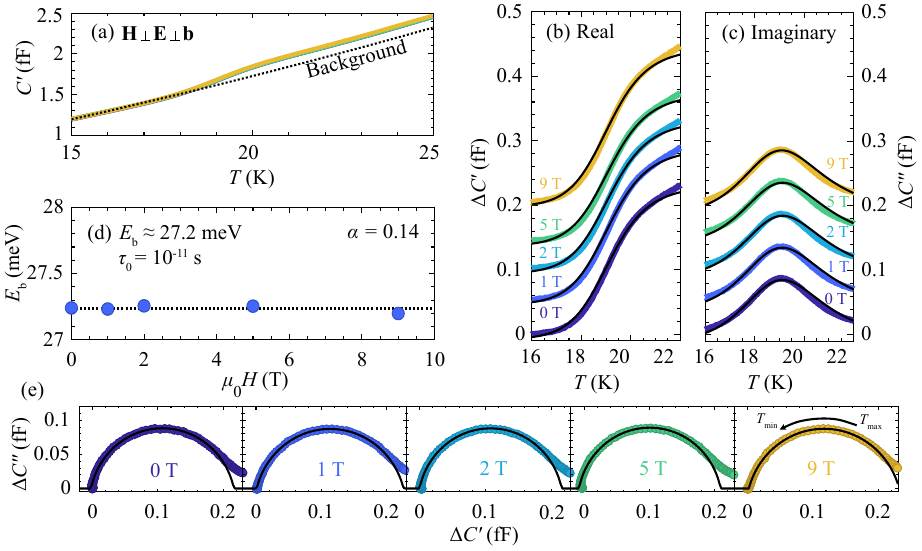}
		\caption{(a) Capacitance measured on coaligned single crystals of \RCMO at several fields as reported in \cite{hayashida2021}.  A phenomenological background ($\Delta C_0$) is determined from the high-temperature capacitance signal.  Data for different fields are perfectly overlapping. (b,c) The reduced real (b) and imaginary (c) components of the signal show good agreement with \eqref{eq:Cole-Cole} as given by the black line.  Measurements for different fields have a vertical offset of 0.05 fF for visibility. (d) Activation energy as a function of field, result of the Cole-Cole fits. The extracted energy gap and attempt time are consistent with optical phonon activity. The values show a negligible field dependence around the average value of 27.2 meV, shown with dashed line.}
		\label{fig:Relaxation_RCMO}
\end{figure*}		

We finally turn to \RCMO and revisit the single crystal capacitance data at high temperature reported in \cite{hayashida2021}. There we have observed a curious dielectric anomaly at much higher temperatures, around 19~K [Fig.2(c) of Ref.~\cite{hayashida2021} and Fig.~\ref{fig:Comparison}]. It involves both a peak in the imaginary part of capacitance and a step in the real component, reminiscent of Debye relaxation.  This is in contrast to the dielectric response of \CCMO, where no dielectric anomalies were found in this temperature range (Fig.~\ref{fig:Comparison}) or up to room temperature (not shown). We note, however, the striking similarity of this plot with the relaxation anomaly seen in the Cs-based compound at much lower temperatures.  

Assuming that in \RCMO we are dealing with a similar relaxation process of intrinsic dipolar degrees of freedom, we analyze these data using the Cole-Cole relaxation model in \eqref{eq:Cole-Cole}.  At such high temperatures the lattice expansion is no longer negligible and contributes to a significant variation on $C$. To account for this effect, a field-independent linear background is subtracted from the data as shown in Fig.~\ref{fig:Relaxation_RCMO}(a).  The resulting complex capacitance components ($\Delta C^{\prime}$ and $\Delta C^{\prime\prime}$) are displayed in Fig.~\ref{fig:Relaxation_RCMO}(b), \ref{fig:Relaxation_RCMO}(c), and \ref{fig:Relaxation_RCMO}(e).


Cole-Cole plots in Fig.~\ref{fig:Relaxation_RCMO}(e) clearly show the characteristic semicircle response of a relaxation process for all values of magnetic field. Note that data are collected at a constant frequency, and the plots show the temperature variation of complex capacitance. A value of $\alpha=0.14$ is extracted from fitting the data and is used in the subsequent analysis.  A good fit to \eqref{eq:Cole-Cole} with activated behavior is obtained with a single $\alpha$, a flat $\Delta C_0 $ = const., and a single attempt time $\tau_0 \approx 10^{-11}$ s (or a frequency of 0.1~THz), shown as black lines in Fig.\ref{fig:Relaxation_RCMO}(b) and \ref{fig:Relaxation_RCMO}(c). The characteristic energy barrier is optimized for each value of magnetic field, showing a negligible field dependence around an average value of $E_b/k_B = 27.2$~meV, as illustrated in Fig.\ref{fig:Relaxation_RCMO}(d).  As opposed to \CCMO, relaxation here is occurring over a much higher energy barrier, and with a much higher attempt rate.  

	\section{Discussion}	

From the above it becomes clear that the phase-boundary anomalies in \CCMO and \RCMO are, essentially,  {\em identical}. The same applies to quantum criticality found at the saturation field, constituting both fantastic instances of 3D QCP BEC of magnons.  

That the relaxation anomaly in \CCMO is something entirely distinct and independent of ferroelectric order is apparent from its lack of response to a DC electric field. Such behavior may be consistent with defects being the origin of the low-energy electric dipolar degrees of freedom. A crude estimate of the energy level shift induced by the DC electric field can be given as $\Delta \sim Zea_0E\approx 25$ mK, assuming the maximum field used in Fig.~\ref{fig:BiasField} of $E =$ 250 kV/m, $Z\sim$1 and a displacement $a_0$ of 0.1 \AA~as typically found in ferroelectric materials \cite{BaTiO1993Structure,BaTiO1998Structure,FerroelectricsDisplacements}. This shift is well over an order of magnitude smaller than the energy barrier mediating relaxation responsible for the anomalies in Fig.~\ref{fig:BiasField}, of the order of $E_b\sim$ 600-800 mK. 

Let us now discuss the potential implications of the qualitative difference between dielectric relaxation in the Cs- and Rb-based compounds. In the latter compound, dipolar degrees of freedom appear to also be present as evidenced in Fig.~\ref{fig:Relaxation_RCMO}. We may speculate that they are the result of crystallographic defects or oxygen vacancies.  However, in \RCMO these electric dipoles relax over a much larger barrier via a different mechanism. As we have previously shown, the 19~K feature in \RCMO is not affected by the magnetic field.  The relaxation is thus unlikely to be of a magnetic origin.  The effects of trapped charge carriers or polarons \cite{PhysRevB.48.13691,PhysRevB.70.174306,Bidault1995Polaronic} can also be ruled out in such a crystalline stoichiometric insulator. A more likely mechanism is one involving an optical phonon. Low temperature dielectric relaxation features like those in Fig.~\ref{fig:Relaxation_RCMO} have been observed in SrTiO$_3$ as a collection of dielectric and ultrasound anomalies between 8 K and 67 K \cite{ang1999, scott_1999}.  Relaxation with an attempt time of 10$^{-12}$ s was found and associated with an optical phonon at 33 meV in energy.  Identification of that phonon as a "silent" mode that can not directly couple to charge indicated that the coupling must be through a defect, which locally breaks symmetry \cite{scott_1999}. These numbers are fully consistent with our observations for \RCMO, suggesting that the relaxation mechanism in this case may likely be due to an optical phonon with a similar coupling.

	\section{Conclusions}
	
The dielectric anomalies associated with the ferroelectric-antiferromagnetic ordering in \CCMO and \RCMO have been shown to be extremely similar, as is the corresponding magnon-BEC behavior at the magnetic field induced quantum critical point. While both materials feature some intrinsic dipolar degrees of freedom, the mechanisms leading to relaxation are very different. In the Cs-compound this happens is via magnons that soften at the magnetic quantum critical point and show electrical activity regardless the ground state.  In the Rb-system relaxation occurs likely via optical phonon modes.  

	\section{Acknowledgments}

	This work was partially supported by the Swiss National Science Foundation, Division II. This work was supported by the Deutsche Forschungsgemeinschaft through ZV 6/2-2, the W\"{u}rzburg-Dresden Cluster of Excellence on Complexity and Topology in Quantum Matter - ct.qmat (EXC 2147, Project No. 390858490) and the SFB 1143 (Project No. 247310070), as well as by HLD at HZDR, member of the European Magnetic Field Laboratory (EMFL).  PAV acknowledges support by the University of Connecticut OVPR Quantum CT seed grant.
	
\bibliography{PRB24_main.bib}
\end{document}